\documentclass[manuscript]{aastex}

\usepackage{graphics,color}
\usepackage{natbib}
\definecolor{dark}{gray}{0.5}
\definecolor{red}{rgb}{1,0,0}
\definecolor{green}{rgb}{0,1,0}
\definecolor{blue}{rgb}{0,0,1}

\shorttitle{The Structure Parameters Of The Milky Way From 2MASS SC}
\shortauthors{Chang et al.}
\begin{document}
\title{The Information Of The Milky Way From 2MASS Whole Sky Star Count: The Structure Parameters}


\author{Chan-Kao Chang\altaffilmark{1}}
\affil{Institute of Astronomy, National Central University, Jhongli, Taiwan}

\author{Chung-Ming Ko\altaffilmark{2}}
\affil{Institute of Astronomy, Department of Physics and Center of Complex Systems, \\
National Central University, Jhongli, Taiwan\\
Department of Physics, University of Hong Kong, Pokfulam Road, Hong Kong}

 \and
\author{Ting-Hung Peng}
\affil{Institute of Astronomy, National Central University, Jhongli, Taiwan}

\altaffiltext{1}{rex@astro.ncu.edu.tw}
\altaffiltext{2}{cmko@astro.ncu.edu.tw}

\begin{abstract}
The $K_s$ band differential star count of the Two Micron All Sky Survey (2MASS)
is used to derive the global structure parameters of the smooth components of
the Milky Way. To avoid complication introduced by other fine structures and
significant extinction near and at the Galactic plane, we only consider
Galactic latitude $|b|>30^{\circ}$ data. The star count data is fitted with a
three-component model: double exponential thin disk and thick disk, and a power
law decay oblate halo. Using maximum likelihood the best-fit local density of
the thin disk is $n_0=0.030\pm 0.002$ stars/pc$^3$. The best-fit scale-height and
length of the thin disk are $H_{z1}=360\pm 10$ pc and $H_{r1}=3.7\pm 1.0$ kpc,
and those of the thick disk are $H_{z2}=1020\pm 30$ pc and $H_{r2}=5.0\pm
1.0$ kpc, the local thick-to-thin disk density ratio is $f_2=7\pm 1$\%. The
best-fit axis ratio, power law index and local density ratio of the oblate halo
are $\kappa=0.55\pm 0.15$, $p=2.6\pm 0.6$ and $f_h=0.20\pm 0.10$\%,
respectively. Moreover, we find some degeneracy among the key parameters (e.g.,
$n_0, H_{z1}, f_2$ and $H_{z2}$). Any pair of these parameters are
anti-correlated to each other. The 2MASS data can be well-fitted by several
possible combinations of these parameters. This is probably the reason that there is
a wide range of values for the structure parameters in literature similar to
this study. Since only medium and high Galactic latitude data are analyzed, the
fitting is insensitive to the scale-lengths of the disks.
\end{abstract}


\section{Introduction}
In the eighteenth century, the famous astronomer William Herschel showed us the
powerful method of star count to understand our own Milky Way \citep{Hers1785}.
The technique has been used since then by generations of astronomers. With
great improvement on data collection over the years, more and more details of
our Milky Way were unfolded. However, the characteristic scales of smooth
Galactic structures (i.e., disks and halo) obtained by previous studies does
not converge to a common value as an outcome of improving data collection
(see Table~\ref{table_ref}). The spread of values is attributed to the
degeneracy of Galactic model parameters (i.e., same star count data could be
fitted equally well by different Galactic models) \citep{Chen2001,
Siegel2002, Juric2008, Bilir2008}.
This is due to the different sky regions
and limiting magnitudes (i.e., limiting volumes) used in these studies
\citep{Siegel2002, Karaali2004, Bilir2006a, Bilir2006b, Juric2008}. On the
contrary, \citet{Bilir2008, Yaz2010} did not show such degeneracy in determining
the Galactic model parameters. This controversy is still
actively debated.
Therefore, systematic all sky surveys with deeper limiting magnitude and wider sky
region, such as the Two Micron All Sky Survey \citep[2MASS;][]{Skrutskie2006},
the Sloan Digital Sky Survey \citep[SDSS;][]{York2000}, the Panoramic Survey
Telescope \& Rapid Response System \citep[Pan-Starrs;][]{Kaiser2002} and the
GAIA mission \citep{Perryman2001} could provide a good opportunity for us to
study our Galaxy from a global
perspective. Free from limited sky
fields, astronomers can acquire many more information from the stellar distribution
of these surveys.

On Galactic structure study, besides the simple and smooth
two-component model \citep{Bahcall1980} or the three-component model
\citep{Gilmore1985}, many more structures have been discovered, such as
inner bars in the Galactic center
\citep{Alves2000, Hammersley2000, vanLoon2003, Nishiyama2005, Cabrera2008},
and flares and warps \citep{Lopez2002, Robin2003, Momany2006, Reyle2009},
which has been contributed the variation of disk model parameters with
Galactic longitude \citep{Cabrera2007, Bilir2008, Yaz2010}.
Moreover, the overdensities in the halo, such as Sagittarius
\citep{Majewski2003}, Triangulum-Andromeda \citep{Rocha2004, Majewski2004},
Virgo \citep{Juric2008}, and in the outer disk, such as Canis Major
\citep{Martin2004}, Monoceros \citep{Newberg2002,Rocha2003}, show the
complexity of the Milky Way. The formation history of our Galaxy is more
complicated than what we thought previously.

On stellar luminosity function
study, a recent study by \citet{Chang2010} using the $K_s$ band star count of
2MASS point source catalog \citep[2MASS PSC;][]{Cutri2003} verified for
the first time the universality hypothesis of the luminosity function (i.e.,
the common practice of assuming one luminosity function for the entire Milky
Way).

We are interested in the global smooth structure of our Galaxy. In view of the
existing fine structures (e.g., flares, warps, overdensities\dots etc.), the
structure of the smooth components can be determined either by including these
fine structures in a grand Galaxy model or by avoiding the sky area
``contaminated'' with these features. The first method demands a complex model,
which involves many more structure parameters, and needs high computing power
to accomplish the fitting task. \citet{Lopez2002} is a good example of this
method using 2MASS data. The second method is clearly simpler but needs
justifications. We observe that (1) the fine structures (e.g., inner bars,
flares and warps), which have observable contribution on 2MASS star count data,
are all confined in the Galactic plane region.
(2) The overdensities or substructures in the outer disk region and the halo are
difficult to identify in general.
Their contribution is negligible in the 2MASS star count data.
Here we quote \citet{Majewski2004} on halo substructure:
``This substructure is typically subtle and obscured by a substantial
foreground veil of disk stars, eliciting its presence requires strategies that
optimize the substructure signal compared to the foreground noise.''

We prefer the second method in this paper and only use Galactic latitude
$|b|>30^{\circ}$ 2MASS $K_s$ band star
count data to obtain the structure parameters of a three-component model.
In addition, the influence of the near infrared extinction in these
regions is small. This also allows us to use a simpler extinction model for
correction. We describe our model in section 2 and the analysis method in
section 3. Section 4 provides the results and a discussion.

\section{The Milky Way Model}
We adopt a three-component model for the smooth stellar distribution of the
Milky Way. It comprises a thin disk, a thick disk and an oblate halo
\citep{Bahcall1980, Gilmore1983}. The total stellar density $n(R,Z)$
at a location $(R,Z)$
is the sum of the thin disk $D_1$, the think disk $D_2$ and the
halo $H$,
\begin{equation}\label{density}
  n(R,Z)=n_0\left[D_1(R,Z)+D_2(R,Z)+H(R,Z)\right]\,,
\end{equation}
where $R$ is the galactocentric distance on the Galactic plane, $Z$ is the
distance from the the Galactic mid-plane and $n_0$ is the local stellar
density of the thin disk at the solar neighborhood.

The stellar distribution of the thin disk $D_1$ and the thick disk $D_2$
decreases exponentially along $R$ and $Z$ (the so called double exponential
disk),
\begin{equation}\label{Di}
  D_i(R,Z)=f_i\exp\left[-\,{(R-R_\odot)\over H_{ri}}-\,{(|Z|-|Z_\odot|)\over H_{zi}}\right]\,,
\end{equation}
where $(R_\odot,Z_\odot)$ is the location of the Sun, $H_{ri}$ is the
scale-length, $H_{zi}$ is the scale-height, and $f_i$ is the density ratio to
the thin disk at the solar neighborhood. The subscript $i=1$ stands for the
thin disk (thus $f_1=1$) and $i=2$ for thick disk. We adopted $R_\odot=8$ kpc
in our model \citep{Reid1993}.

The halo is a power law decay oblate spheroid flattening in the $Z$ direction,
\begin{equation}\label{SH}
  H(R,Z)=f_h\left[R^2+(Z/\kappa)^2\over R_\odot^2+(Z_\odot/\kappa)^2\right]^{-p/2}\,,
\end{equation}
where $\kappa$ is the axis ratio, $p$ is the power index and $f_h$ is the local
halo-to-thin disk density ratio.

\citet{Chang2010} showed that the whole sky $K_s$ band luminosity function can
be well approximated by a single power law with a power law index
$\gamma=1.85\pm0.035$, a bright cutoff at $M_b=-7.86\pm0.60$ and a faint cutoff
at $M_f=6.88\pm0.66$. We adopt this luminosity function in the following analysis.
In normalized form, it is
\begin{equation}\label{LF}
  \psi(M_{K_s})={2\log_{\rm e}10\ (\gamma-1)\over
  5\left[10^{2(\gamma-1)M_{f}/5}-10^{2(\gamma-1)M_{b}/5}\right]}\,10^{2(\gamma-1)M/5}\,.
\end{equation}
Note that $\psi(M)$ includes all luminosity classes.
In our analysis the observing magnitude range is $5 \le K_s \le 14$ mag.
The corresponding distances
of bright cutoff, faint cutoff and $M_{K_s}=0$ are 3.4 kpc to 216 kpc, 4 pc to
265 pc and 0.1 kpc to 6.3 kpc, respectively.

Although we only use NIR data in the medium and high Galactic latitude regions,
we still need to correct possible interstellar extinction. We adopt the new
COBE/IRAS result \citep{Chen1999} and convert it to the $K_s$ band extinction
by $A_{K_s}/E(B-V)=0.367$ \citep{Schlegel1998}. This extinction model is then
applied to our simulation data. The extinction values of most of our analyzed
regions are $A_{K_s}< 0.03$.

\section{The Data and Analysis Method}
\subsection{The 2MASS Data}
The 2MASS Point Source Catalog \citep[2MASS PSC,][]{Cutri2003} is employed to
carry out the $K_s$ band differential star count for the entire Milky Way. We
divide the whole sky into 8192 nodes according to level 5 Hierarchical
Triangular Mesh \citep[HTM,][]{Kunszt2001}. The level 5 HTM samples the whole
sky in roughly equal area with an average angular distance about 2 degrees
between any two neighboring nodes. The amount of stars within 1 degree radius
of each node (i.e., each node covers $\pi$ square degree) is then retrieved with
a bin size $K_s$=0.5 mag via 2MASS online data service
\citep[][catalog]{Cutri2003}. Our selection criterion is: the object must be
detected in all $J, H, K_s$ bands and has signal-to-noise ratio $\ge 5$. Since
the limiting magnitude of 2MASS $K_s$ band is 14.3 mag, which has 10
signal-to-noise ratio and 99\% completeness \citep[see Table~1
in][]{Skrutskie2006}, and $K_s \le 5$ mag objects have relatively large
photometric error, we only compare 2MASS data with our simulation data from
$K_s=5$ to $14$ mag.

In order to minimize the effects coming from the close-to-Galactic-plane fine
structures (e.g., flares, warps, arms, budge and bars\dots etc., which have
considerable contribution to the star count data), and the relatively complex
extinction correction at the low Galactic latitude region,
we avoid low galactic latitudes, and only consider data in Galactic
latitude $|b|>30^{\circ}$. Although several
overdensities in the halo (such as Sagittarius, Triangulum-Andromeda,
Virgo\dots etc.) were identified, they cannot be
picked up
from the
overwhelming foreground field stars on star count data without additional
information \citep[e.g., color, distance, metallicity\dots
etc.,][]{Majewski2004}. Therefore, their contribution to the 2MASS $K_s$ band
differential star count is negligible and will not affect our result. We also
exclude the areas around Large and Small Magellanic Clouds for their
significant stellar population.

\subsection{The Analysis Method}\label{method}
The Maximum Likelihood Method (Bienayme et al. 1987) is applied to compare the
$K_s$ band 2MASS differential star counts and the simulation data to search for
the best-fit structure parameters of the three-component Milky Way model. Our
fitting strategy is as follows:
\begin{enumerate}\label{fitting_steps}
\item Take $R_\odot=8$ kpc;
\item choose one $Z_\odot$ and work out the maximum likelihood value by fitting the 9 parameters
$(n_0,H_{z1},H_{r1},f_2,H_{z2},H_{r2},f_h,\kappa,p)$;
\item repeat step 2 for other $Z_\odot$;
\item pick the $Z_\odot$ corresponds to the maximum of the maximum likelihood values in step 3;
\item repeat steps 2 to 4 for finer grid size of the 9-parameter fit and a narrower range of
$Z_\odot$ around the one found in step 4;
\item the uncertainty is estimated by adding Poisson noise on the simulation data to see
how the likelihood varies. The difference of the likelihoods of 500
realizations of the same model, differed by the Poisson statistics only,
gives a range of likelihood around the maximum likelihood that defines the
confidence level.
\end{enumerate}
Table~\ref{para_space} lists our searching parameter space and the finest grid
size we used. The key parameters in our study are $n_0$, $H_{z1}$, $f_2$ and
$H_{z2}$ (see Eqs. (1)-(2)). The first two play a primary role on the variation
of the likelihood value and the latter two play a secondary role. The other
five parameters $H_{r1}$, $H_{r2}$, $f_h$, $\kappa$ and $p$ (see Eqs. (1)-(3))
are non-key parameters, which play a minor role and do not affect the
likelihood value as much as the key parameters.

\section{The Results and Discussion}
Table~\ref{result} lists our best-fit results with the corresponding
uncertainties. Fig.~\ref{2Dcon} shows contour plots of likelihood against
different pairs of parameters. The contour changes dramatically along the key
parameters $n_0$, $H_{z1}$, $f_2$ and $H_{z2}$, but relatively mild along other
parameters $H_{r1}$, $H_{r2}$, $f_h$, $p$ and $\kappa$.
This indicates the importance of the key parameters in determining the best-fit result.

Degeneracies exist between some pairs of key parameters, such as $(n_0,
H_{z1})$, $(n_0, f_2)$, $(H_{z1}, f_2)$\dots etc. Here degeneracy means that
the likelihood value stays almost the same when the pairs of parameters change
together in a particular way. Similar degeneracy between the local
thick-to-thin disk ratio $f_2$ and the scale-height of the thick disk $H_{z2}$
has been reported in \citet{Chen2001,Siegel2002,Juric2008,Bilir2008}.
Consequently, it is possible that different combinations of parameters can be
regarded as `acceptable' fitting. For example, if we choose a higher local
stellar density $n_0$, then we can pick a smaller $H_{z1}$ such that the
likelihood value is very close to the maximum likelihood value and assign it as
the `best-match' scale-height of the thin disk. Therefore, a thin light color
diagonal strip shows on the $n_0$ against $H_{z1}$ contour plot in
Fig.~\ref{2Dcon}. Besides, similar trends happen in other pairs of key
parameters. When one parameter of the pair is higher, we can get a similar
likelihood value by lowering the other parameter (see the corresponding
two-parameter contour plots of key parameters in Fig.~\ref{2Dcon}). This
anti-correlation is not unexpected. The number of stars along the line of sight
in the model increases when any one of the key parameters increases. Thus for a
given observed number of stars, an increase in one key parameter can be
compensated by a decrease in the other. Perhaps this is the reason that our
best-fit scale-height of the thin disk, $H_{z1}=360$ pc, is somewhat larger
than the reported values, $H_{z1}=285$ pc, in \citet{Lopez2002} study (they
also use star count of 2MASS to obtain Galactic model parameters). Our best-fit
local stellar density from $K_s=$-8 to 6.5 mag is $n_0\sim 0.030$ star/pc$^3$,
which is about half of $\sim 0.056$ star/pc$^3$ of the corresponding value
cited in \citet{Eaton1984}, and a bit lower than $\sim 0.032$ star/pc$^3$ of
the corresponding value cited in \citet{Lopez2002}. As a result, our fitting
tends to choose a larger scale-height of the thin disk. If we force our local
stellar density to be comparable with that of \citet{Lopez2002}, then the
corresponding `best-fit' scale-height of the thin disk would be $\sim$320 pc,
which is closer to their result. If we choose even higher local stellar
density, then the `best-fit' scale-height of the thin disk would be made
between 200 to 300 pc, which is similar to the most of recent studies (see
Table~\ref{table_ref}). Moreover, we do not apply binarism correction in our
analysis, and it has been shown that scale-length and scale-height might be
underestimated without binarism correction \citep{Siegel2002, Juric2008,
Ivezic2008, Yaz2010}.

For our purpose, we deem that in order to lift the degeneracy it is crucial to
have a reliable near infrared luminosity function by observation or a near
infrared local stellar density. Unfortunately, a systematic study in this
direction is yet to come. Some related studies, such as synthetic luminosity
function \citep[see e.g.,][]{Girardi2005} or luminosity function transformed
from optical observation \citep[see e.g.,][]{Wainscoat1992} do exist, but some
uncertainties still need to be settled (e.g., the initial mass function,
mass-luminosity relation for NIR and color transformation between different
wavelengths\dots etc.). Once the `true' local stellar density is known, the
`true' structure of the Milky Way would be revealed.

\begin{figure}
  \epsscale{0.8}
  \plotone{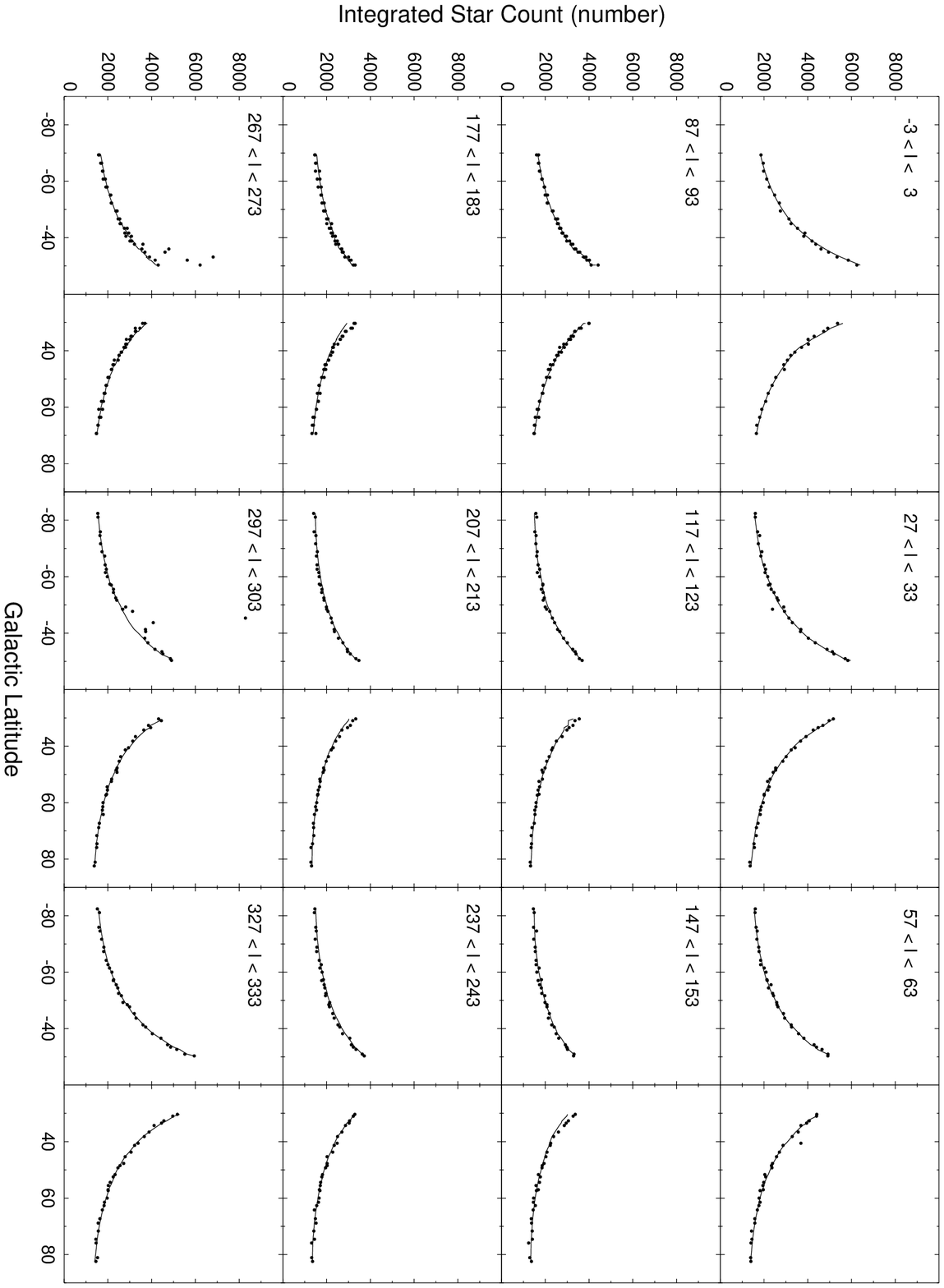}
  \caption{Integrated star count from $K_s=$5  to 14 mag of 2MASS data
  as a function of the Galactic latitude.
  Solid line is the model prediction.}
  \label{surf_b}
  \end{figure}

\begin{figure}
  \epsscale{0.8}
  \plotone{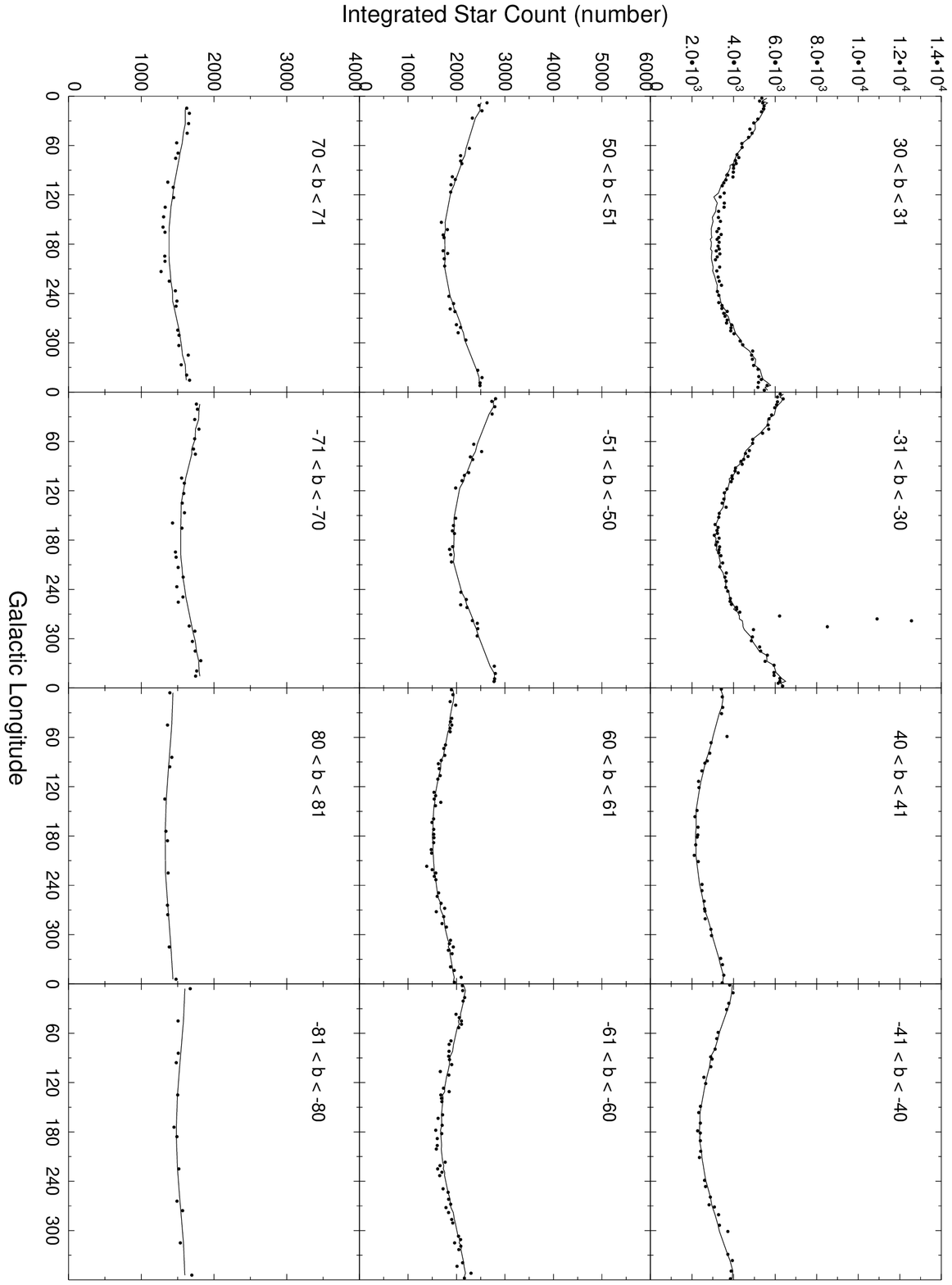}
  \caption{Integrated star count from $K_s=$5  to 14 mag of 2MASS data
  as a function of the Galactic longitude.
  Solid line is the model prediction.}
  \label{surf_l}
  \end{figure}

In order to see how good the agreement between 2MASS data and our best-fit
model, we show an all sky map of the ratios of observed to predicted integrated
star count from $K_s=$5 to 14 mag for each node as a function of position on
the sky in Fig.~\ref{comparison}, and the color indicates the values of the
ratios. We do not see obvious deviation in the Galactic latitude
$|b|>30^{\circ}$ areas, but only in the Large Magellanic Cloud and the Small
Magellanic Cloud areas. For comfirmation, we plot integrated star count from
$K_s=$5 to 14 mag along Galactic longitude and latitude for $|b|>30^{\circ}$
areas in Figs.~\ref{surf_b} \& \ref{surf_l}. We see 2MASS data and our best-fit
model agree well and only some small deviations in the nodes at the
anti-Galactic center $b\sim30^{\circ}$ areas. The significant spikes in
Figs.~\ref{surf_b} \& \ref{surf_l} are due to the populations of the Large
Magellanic Cloud and the Small Magellanic Cloud. For testing how these small
deviations affect the key parameters selection, we exclude these small
deviations and re-analyze the data with fixed non-key parameters. It shows only
the scale-height of thick disk shifts slightly but still within our error
estimation. Besides, we do not see any significant differences of reported
overdensities in sky regions with $|b|>30^{\circ}$. Thus, we conclude that the
three-component model can describe the Milky Way structure sufficiently well
for high Galactic latitude regions and the single power law luminosity function
of \citet{Chang2010} is a good approximation as well. Since our main purpose is
to search for the global Galactic model parameters, we do not try to explore
the best-fit result for each node individually as what \citet{Cabrera2007,
Bilir2008, Yaz2010} have done in their studies to seek the variations of disk
model parameters with Galactic longitude. Instead, we treat the differences of
2MASS data to our best-fit model as deviations from a global smooth
distribution. We believe that similar variations in disk model parameters will
be obtained if we take similar analysis procedure (i.e., searching best-fit
parameters for each node), but this is beyond the scope of this work.
Because we do not consider flares, warps and other overdensities in our
model, there are some discrepancies between 2MASS data and the model in the low
Galactic latitude regions (see Fig.~\ref{comparison}). These fine structures
make the star distribution more fluffy in the vertical direction toward the
edge of the Milky Way. Hence the discrepancy between 2MASS data and the model
increases vertically towards the anti-Galactic center region. In addition, the
absence of Galactic bulge in our model contributes to the large discrepancy in
Galactic center areas. The difference in the low Galactic latitude region needs
more delicate analysis to rectify \citep[see, e.g.,][]{Lopez2002, Momany2006}.

\subsection{Summary}
In summary, we set forth to study the global smooth structure of the Milky Way
by a three-component stellar distribution model which comprises two double
exponential disks (one thin and one thick) and an oblate halo. The $K_s$ band
2MASS star count is used to determine the structure parameters. To avoid the
complication introduced by the fine structures and complex extinction
correction close to Galactic plane, we use only Galactic latitude
$|b|>30^{\circ}$ data. There are 10 parameters in the model, but only four of
them play the dominant  role in the fitting process. They are the local stellar
density of the thin disk $n_0$, the local density ratio of thick-to-thin disk
$f_2$, and the scale-height of the thin and thick disks $H_{z1}$ and $H_{z2}$.
The best-fit result is listed in Table~\ref{result}. In short the scale-height
of the thin and thick disks are $360\pm 10$ pc and $1020\pm 30$ pc,
respectively; the scale-length of the thin disk is $3.7\pm 1.0$ kpc and that of
thick disk is $5.0\pm 1.0$ kpc (the uncertainty in scale-length is large
because it is not very sensitive to high latitude data.) The local stellar
density ratio of thick-to-thin disk and halo-to-thin disk are $7\pm 1$\% and
$0.20\pm 0.10$\%, respectively. The local stellar density of the thin disk is
$0.030\pm 0.002$ stars/pc$^3$.

\begin{figure}
  \epsscale{0.8}
  \plotone{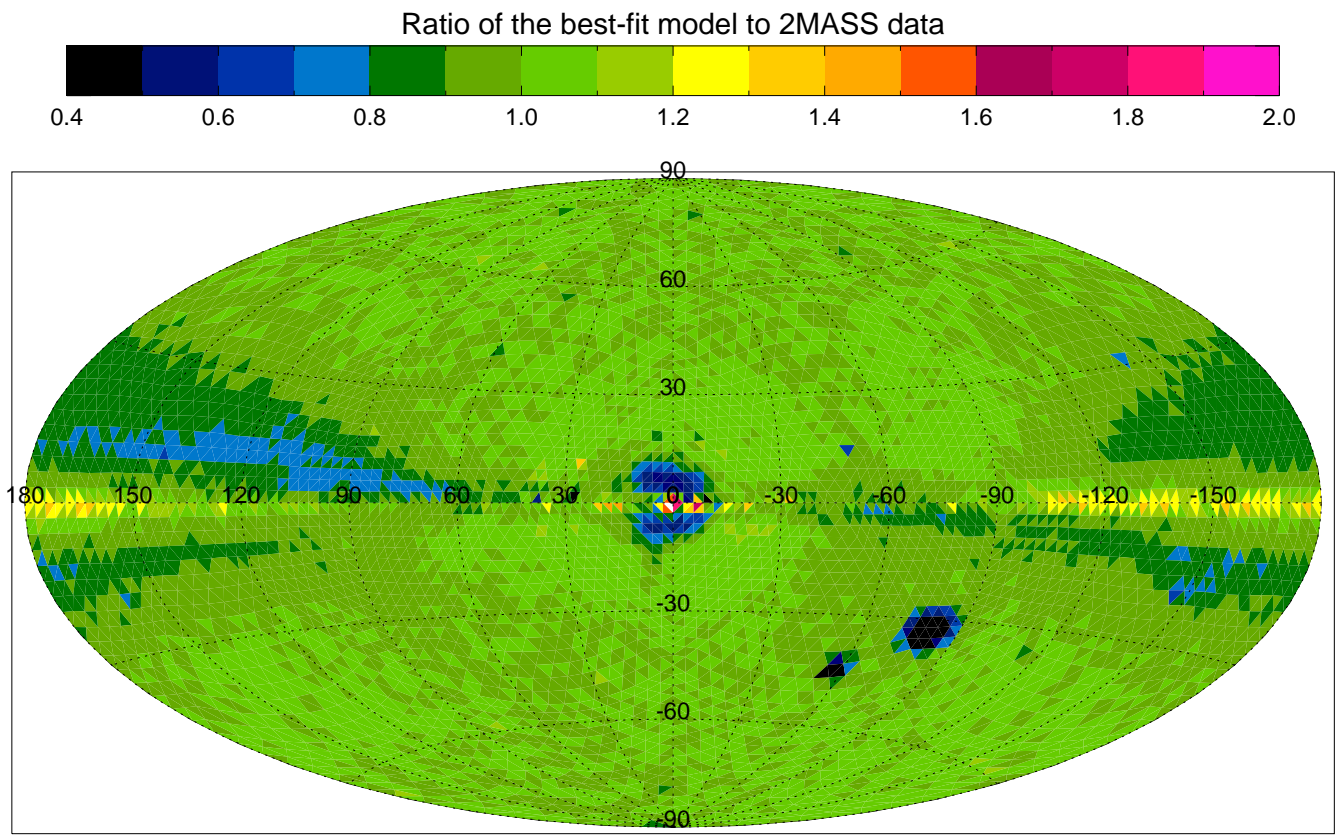}
  \caption{The ratios of observed to model integrated star count from $K_s=$5
  to 14 mag.}
  \label{comparison}
  \end{figure}

An all sky comparison of the 2MASS data to our best-fit model is shown in
Fig.~\ref{comparison}. A good agreement in the Galactic latitude
$|b|>30^{\circ}$ areas is expected from our fitting procedure. In low Galactic
latitude regions, fine structures (such as flares, warps\dots etc.) increase
the effective scale-height towards the edge of the Milky way. This is reflected
in the fan-like increase in discrepancy towards the anti-Galactic center
regions.

\begin{figure}
  \epsscale{0.8}
  \plotone{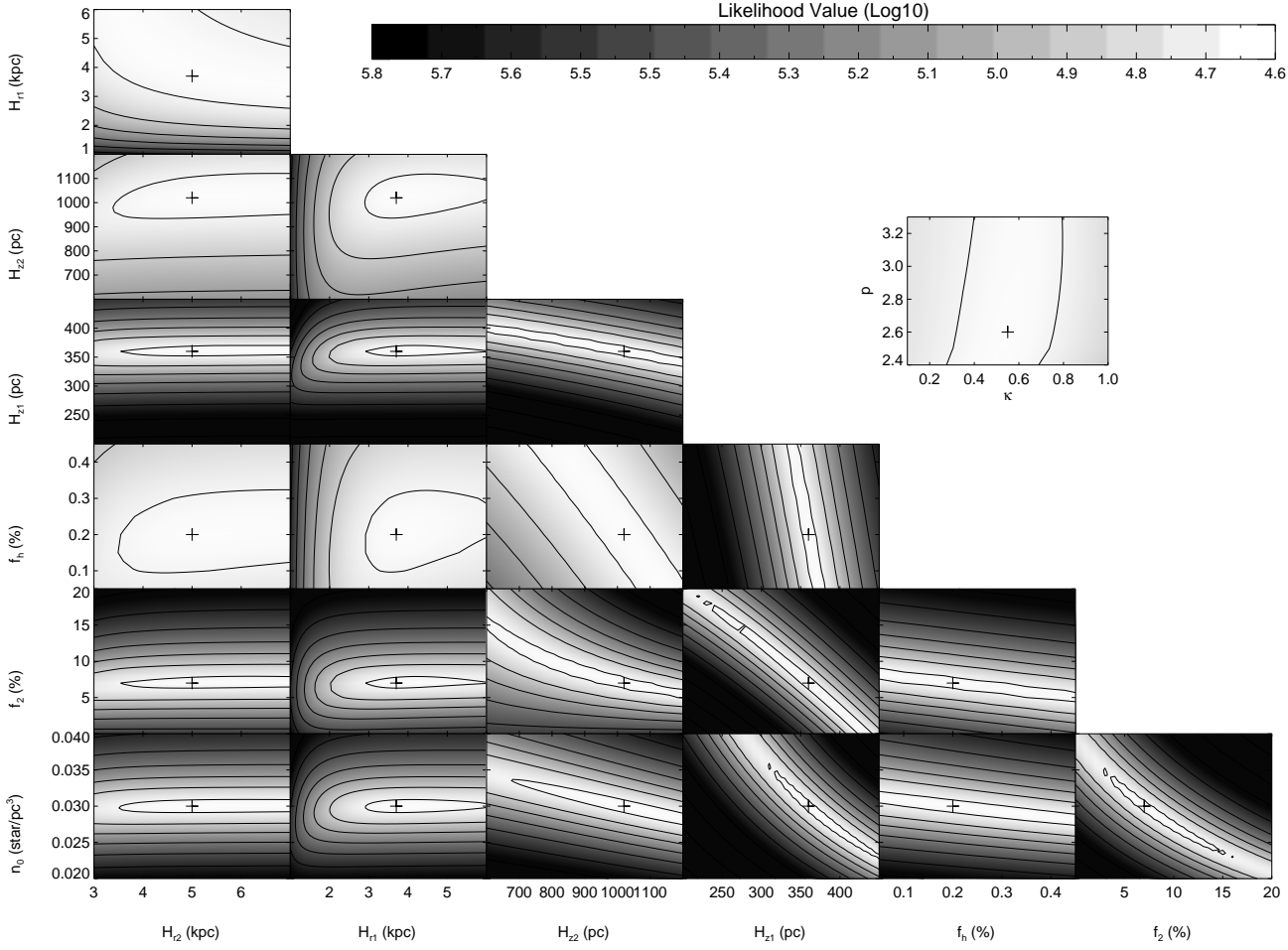}
  \caption{Contour plots of likelihood value of pairs of parameters
  in the Milky Way model. $n_0$, $f_2$, $H_{z1}$, $H_{z2}$ are key parameters of the model.
  In this plot $R_\odot=8$ kpc and $Z_\odot=25$ pc. The cross marks our best-fit values.
  The innermost contour is at logarithmically maximum likelihood value 4.67,
  and the rest are logarithmically spaced in steps of 0.2 dex.}
  \label{2Dcon}
  \end{figure}

Degeneracy (i.e., different combinations of parameters give similar likelihood
values) is found in pairs of key parameters
(see Fig.~\ref{2Dcon}). Thus different combinations of parameters may fit the
data almost as good as the best-fit one, and these are all legitimate
`acceptable' fitting in view of the uncertainty. Therefore, accompanying our
lower local stellar density $0.030$ stars/pc$^3$ (from $K_s=-$8 to 6.5 mag) is
a higher thin disk scale-height $360$ pc. In the context of NIR star count, the
NIR luminosity function or the NIR local stellar density is imperative to
determine the scale-height and other Milky Way structure parameters. We hope
that systematic study on the luminosity function and the local stellar density
in near infrared will be available in the near future.

\acknowledgements
We acknowledge the use of the Two Micron All Sky Survey Point
Source Catalog (2MASS PSC).
We would like to thank the anonymous referee, whose advice greatly improves the
paper.
CMK is grateful to S. Kwok and K.S. Cheng for their
hospitality during his stay at the Department of Physics, Faculty of Science,
University of Hong Kong. This work is supported in part by the National Science
Council of Taiwan under the grants NSC-98-2923-M-008-001-MY3 and
NSC-99-2112-M-008-015-MY3.

\begin{deluxetable}{lllllllll}
\tabletypesize{\tiny}  \setlength{\tabcolsep}{0.03in} \tablecaption{Previous
Galactic models. The parentheses are the corrected values for binarism. The
asterisk denotes the power-law index replacing Re. References indicating with
the original result table mean Galactic longitude or limiting magnitude
dependent Galactic model parameters. \label{table_ref}} \tablewidth{0pt}
\tablehead{ \colhead{$H_{z1}$ (pc)} & \colhead{$H_{r1}$ (kpc)} & \colhead{$f_2$
(\%)} & \colhead{$H_{z2}$ (kpc)} & \colhead{$H_{r2}$ (kpc)} &
\colhead{$f_h$(\%)} & \colhead{Re(S) (kpc)} & \colhead{$\kappa$} &
\colhead{Reference}} \startdata
310 - 325   &    -  &   0.0125-0.025    &   1.92 - 2.39 &    -  &    -  &    -  &    -  &   \citet{Yoshii1982}  \\
300 &    -  &   0.02    &   1.45    &    -  &   -   &   -   &   -    &   \citet{Gilmore1983} \\
325 &    -  &   0.02    &   1.3     &    -  &   0.002   &   3   &   0.85    &   \citet{Gilmore1984} \\
280 &    -  &   0.0028  &   1.9 &    -  &   0.0012  &    -  &    -  &   \citet{Tritton1984} \\
125-475 &    -  &   0.016   &   1.18 - 2.21 &    -  &   0.0013  &    3.1*  &   0.8 &   \citet{Robin1986}   \\
300 &    -  &   0.02    &   1   &    -  &   0.001   &    -  &   0.85    &   \citet{delRio1987}  \\
285 &    -  &   0.015   &   1.3 - 1.5   &    -  &   0.002   &   2.36    &   Flat    &   \citet{Fenkart1987} \\
325 &    -  &   0.0224  &   0.95    &    -  &   0.001   &   2.9 &   0.9 &   \citet{Yoshii1987}  \\
249 &    -  &   0.041   &   1   &    -  &   0.002   &   3   &   0.85    &   \citet{Kuijken1989} \\
350 &   3.8 &   0.019   &   0.9 &   3.8 &   0.0011  &   2.7 &   0.84    &   \citet{Yamagata1992}    \\
290 &    -  &    -  &   0.86    &    -  &    -  &   4   &    -  &   \citet{vonHippel1993}   \\
325 &    -  &   0.020-0.025  &   1.6-1.4 &    -  &   0.0015  &   2.67 &   0.8 &   \citet{ReidMajewski1993}    \\
325 &   3.2 &   0.019   &   0.98    &   4.3 &   0.0024  &   3.3 &   0.48    &   \citet{Larsen1996}  \\
250-270 &   2.5 &   0.056   &   0.76    &   2.8 &   0.0015  &   2.44 - 2.75*    &   0.60 - 0.85 &   \citet{Robin1996, Robin2000}    \\
260 &   2.3 &   0.074   &   0.76    &   3 &    -  &    -  &    -  &   \citet{Ojha1996}    \\
290 &   4   &   0.059   &   0.91    &   3   &   0.0005  &   2.69    &   0.84    &   \citet{Buser1998, Buser1999}    \\
240 &   - &   0.061   &   0.79    &   - &    -  &    -  &    -  &   \citet{Ojha1999}    \\
280/267  &   -    &   0.02    &   1.26/1.29   &    -  &    -  &   2.99*   &   0.63    &   \citet{Phleps2000}  \\
330 &   2.25    &   0.065 - 0.13    &   0.58 - 0.75 &   3.5 &   0.0013  &    -  &   0.55    &   \citet{Chen2001}    \\
-   &   2.8 &   3.5     &   0.86        &   3.7     & -     & -         & -     & \citet{Ojha2001} \\
280(350)    &   2 - 2.5 &   0.06 - 0.10 &   0.7 - 1.0 (0.9 - 1.2)   &   3 - 4  &   0.0015  &    -  &   0.50 - 0.70 &   \citet{Siegel2002}  \\
285 &   1.97    &    -  &    -  &    -  &    -  &    -  &    -  &   \citet[][]{Lopez2002}   \\
 -  &   3.5 &   0.02-0.03   &   0.9 &   4.7 &   0.002-0.003 &   4.3 &   0.5-0.6 &   \citet{Larsen2003}  \\
320 &    -  &   0.07    &   0.64    &    -  &   0.00125 &    -  &   0.6 &   \citet{Du2003}  \\
265-495 &   -  &   0.052-0.098 &   0.805-0.970  &    -  &   0.0002-0.0015   &    -  &   0.6-0.8 &   \citet[][Table 16]{Karaali2004} \\
268 &   2.1 &   0.11    &   1.06    &   3.04    &    -  &    -  &    -  &   \citet{Cabrera2005} \\
300  &   -    &   0.04-0.10    &   0.9   &    -  &    -  &   3/2.5*   &   1/0.6    &   \citet{Phleps2005}  \\
220 &   1.9 &    -  &    -  &    -  &    -  &    -  &    -  &   \citet{Bilir2006a}  \\
160-360 &    -  &   0.033-0.076 &   0.84-0.87   &    -  &   0.0004-0.0006   &    -  &   0.06-0.08   &   \citet[][Table 15]{Bilir2006b}  \\
301/259  &    -  &   0.087/0.055    &   0.58/0.93    &    -  &   0.001   &    -  &   0.74    &   \citet[][Table 5]{Bilir2006c}   \\
220-320 &    -  &   0.01-0.07   &   0.6-1.1 &    -  &   0.00125 &    -  &   $>$0.4    &   \citet{Du2006}  \\
206/198  &    -  &   0.16/0.10    &   0.49/0.58    &    -  &    -  &    -  &   0.45    &   \citet{Ak2007}  \\
140-269 &    -  &   0.062-0.145   &   0.80-1.16   &    -  &    -  &    -  &    -  &   \citet[][Table 1]{Cabrera2007}   \\
220-360 &   1.65-2.52   &   0.027-0.099   &   0.62-1.03   &   2.3-4.0 &   0.0001-0.0022   &    -  &   0.25-0.85   &   \citet[][Table 3]{Karaali2007} \\
167-200 &    -  &   0.055-0.151 &   0.55-0.72   &    -  &   0.0007-0.0019    &    -  &   0.53-0.76   &   \citet[][Table 1]{Bilir2008}    \\
245(300)    &   2.15(2.6)   &   0.13(0.12)  &   0.743(0.900)    &   3.261(3.600)    &   0.0051  &   2.77*   &   0.64    &   \citet{Juric2008}   \\
325-369 &   1.00-1.68   &   0.0640-0.0659   &   0.860-0.952 &   2.65-5.49   &   0.0033-0.0039   &    -  &   0.0489-0.0654   &   \citet[][Table 1]{Yaz2010}  \\
360 &   3.7 &   0.07    &   1.02    &   5   &   0.002   &   2.6*    &   0.55    &   This Work   \\
\enddata
\end{deluxetable}

\begin{deluxetable}{lcc}
\tabletypesize{\scriptsize} \tablecaption{The searching parameter space.
\label{para_space}} \tablewidth{0pt} \tablehead{ \colhead{} & \colhead{Range} &
\colhead{Grid size}} \startdata
Thin Disk &  \\
\quad $H_{r1}$ & 1.0-6.0 kpc  & 100 pc \\
\quad $H_{z1}$ & 200-450 pc   & 10 pc \\
\quad $n_0$    & 0.02-0.04 stars/pc$^3$  & 0.002 stars/pc$^3$ \\
\quad $Z_\odot$ & 11-35 pc    & 2.5 pc \\
Thick Disk & \\
\quad $H_{r2}$ & 3.0-7.0 kpc & 200 pc \\
\quad $H_{z2}$ & 400-1200 pc & 20 pc \\
\quad $f_2$ & 0-20\% & 1\% \\
Spheroid & \\
\quad $\kappa$ & 0.1-1.0  & 0.05  \\
\quad $p$ & 2.3-3.3 & 0.1  \\
\quad $f_h$ & 0-0.45\% & 0.05\% \\
\enddata
\end{deluxetable}

\begin{deluxetable}{lcc}
\tabletypesize{\scriptsize} \tablecaption{The best-fit Milky Way model.
\label{result}} \tablewidth{0pt} \tablehead{ \colhead{} & \colhead{Value} &
\colhead{Uncertainty}} \startdata
Thin Disk &  \\
\quad $H_{r1}$ & 3.7 kpc &  1.0 kpc \\
\quad $H_{z1}$ & 360 pc  &  10 pc   \\
\quad $n_0$    & 0.030 stars/pc$^3$    & 0.002 stars/pc$^3$    \\
\quad $Z_\odot$ & 25 pc   & 5 pc \\
Thick Disk & \\
\quad $H_{r2}$ & 5.0 kpc & 1.0 kpc \\
\quad $H_{z2}$ & 1020 pc & 30 pc \\
\quad $f_2$    & 7\%     & 1 \% \\
Spheroid & \\
\quad $\kappa$ & 0.55        & 0.15  \\
\quad $p$   & 2.6        & 0.6  \\
\quad $f_h$ & 0.20\%     & 0.10\% \\
\enddata
\end{deluxetable}

\clearpage

\begin{thebibliography}{}
\bibitem[Ak et al.(2007)]{Ak2007} Ak, S., Bilir, S., Karaali, S., \& Buser, R.\ 2007, AN, 328, 169
\bibitem[Alves(2000)]{Alves2000} Alves, D.R.\ 2000, \apj, 539, 732
\bibitem[Bahcall \& Soneira(1980)]{Bahcall1980} Bahcall, J.N., \& Soneira, R.M.\ 1980, \apjl, 238, 17
\bibitem[Bienayme et al.(1987)]{Bienayme1987} Bienayme, O., Robin, A.C., \& Creze, M.\ 1987, \aap, 180, 94
\bibitem[Bilir et al.(2006a)]{Bilir2006a} Bilir, S., Karaali, S., Ak, S., Yaz, E., \& Hamzao{\u g}lu, E.\ 2006, \na, 12, 234 (2006a)
\bibitem[Bilir et al.(2006b)]{Bilir2006b} Bilir, S., Karaali, S., \& Gilmore, G.\ 2006, \mnras, 366, 1295 (2006b)
\bibitem[Bilir et al.(2006c)]{Bilir2006c} Bilir, S., Karaali, S., G{\"u}ver, T., Karata{\c s}, Y., \& Ak, S.~G.\ 2006, AN, 327, 72 (2006c)
\bibitem[Bilir et al.(2008)]{Bilir2008} Bilir, S., Cabrera-Lavers, A., Karaali, S., Ak, S., Yaz, E., \& L{\'o}pez-Corredoira, M.\ 2008, \pasa, 25, 69
\bibitem[Buser et al.(1998)]{Buser1998} Buser, R., Rong, J., \& Karaali, S.\ 1998, \aap, 331, 934
\bibitem[Buser et al.(1999)]{Buser1999} Buser, R., Rong, J., \& Karaali, S.\ 1999, \aap, 348, 98
\bibitem[Cabrera-Lavers et al.(2005)]{Cabrera2005} Cabrera-Lavers, A., Garz{\'o}n, F., \& Hammersley, P.~L.\ 2005, \aap, 433, 173
\bibitem[Cabrera-Lavers et al.(2007)]{Cabrera2007} Cabrera-Lavers, A., Bilir, S., Ak, S., Yaz, E., \& L{\'o}pez-Corredoira, M.\ 2007, \aap, 464, 565
\bibitem[Cabrera-Lavers et al.(2008)]{Cabrera2008} Cabrera-Lavers, A., Gonz{\'a}lez-Fern{\'a}ndez, C., Garz{\'o}n, F., Hammersley, P.~L., \& L{\'o}pez-Corredoira, M. \ 2008, \aap, 491, 781
\bibitem[Chang et al.(2010)]{Chang2010} Chang, C.-K., Ko, C.-M., \& Peng, T.-H.\ 2010, \apj, 724, 182
\bibitem[Chen et al.(1999)]{Chen1999} Chen, B., Figueras, F., Torra, J., Jordi, C., Luri, X., \& Galadi-Enriquez, D.\ 1999, \aap, 352, 459
\bibitem[Chen et al.(2001)]{Chen2001} Chen, B., et al. (the SDSS collaboration)\ 2001, \apj, 553, 184
\bibitem[Cutri et al.(2003)]{Cutri2003} Cutri, R.M., et al.\ 2003, The IRSA 2MASS All-Sky Point Source Catalog, NASA/IPAC Infrared Science Archive. http://irsa.ipac.caltech.edu/applications/Gator/
\bibitem[del Rio \& Fenkart(1987)]{delRio1987} del Rio, G., \& Fenkart, R.\ 1987, \aaps, 68, 397
\bibitem[Du et al.(2003)]{Du2003} Du, C., et al.\ 2003, \aap, 407, 541
\bibitem[Du et al.(2006)]{Du2006} Du, C., Ma, J., Wu, Z., \& Zhou, X.\ 2006, \mnras, 372, 1304
\bibitem[Eaton et al.(1984)]{Eaton1984} Eaton, N., Adams, D.J., \& Giles, A.B.\ 1984, \mnras, 208, 241
\bibitem[Fenkart \& Karaali(1987)]{Fenkart1987} Fenkart, R., \& Karaali, S.\ 1987, \aaps, 69, 33
\bibitem[Gilmore(1984)]{Gilmore1984} Gilmore, G.\ 1984, \mnras, 207, 223
\bibitem[Gilmore \& Reid(1983)]{Gilmore1983} Gilmore, G., \& Reid, N.\ 1983, \mnras, 202, 1025
\bibitem[Gilmore \& Wyse(1985)]{Gilmore1985} Gilmore, G., \& Wyse, R.~F.~G.\ 1985, \aj, 90, 2015
\bibitem[Girardi et al.(2005)]{Girardi2005} Girardi, L., Groenewegen, M.A.T., Hatziminaoglou, E., \& da Costa, L.\ 2005, \aap, 436, 895
\bibitem[Hammersley et al.(2000)]{Hammersley2000} Hammersley, P.L., Garz{\'o}n, F., Mahoney, T.J., L{\'o}pez-Corredoira, M., \& Torres, M.A.P.\ 2000, \mnras, 317, L45
\bibitem[Herschel(1785)]{Hers1785} Herschel, W.\ 1785, Royal Society of London Philosophical Transactions Series I, 75, 213
\bibitem[Ivezi{\'c} et al.(2008)]{Ivezic2008} Ivezi{\'c}, {\v Z}., et al.\ 2008, \apj, 684, 287
\bibitem[Juri{\'c} et al.(2008)]{Juric2008} Juri{\'c}, M., et al.\ 2008, \apj, 673, 864
\bibitem[Kaiser et al.(2002)]{Kaiser2002} Kaiser, N., et al.\ 2002, \procspie, 4836, 154
\bibitem[Karaali et al.(2004)]{Karaali2004} Karaali, S., Bilir, S., \& Hamzao{\u g}lu, E.\ 2004, \mnras, 355, 307
\bibitem[Karaali et al.(2007)]{Karaali2007} Karaali, S., Bilir, S., Yaz, E., Hamzao{\u g}lu, E., \& Buser, R.\ 2007, \pasa, 24, 208
\bibitem[Kuijken \& Gilmore(1989)]{Kuijken1989} Kuijken, K., \& Gilmore, G.\ 1989, \mnras, 239, 605
\bibitem[Kunszt et al.(2001)]{Kunszt2001}Kunszt, P.Z., Szalay A.S., \& Thakar A.R.\ 2001, Proc. of the MPA/ESO/MPE workshop (Springer-Verlag Berlin Heidelberg), pp. 631
\bibitem[Larsen(1996)]{Larsen1996} Larsen, J.~A.\ 1996, Ph.D.~Thesis,
\bibitem[Larsen \& Humphreys(2003)]{Larsen2003} Larsen, J.~A., \& Humphreys, R.~M.\ 2003, \aj, 125, 1958
\bibitem[Lopez-Corredoira et al.(2002)]{Lopez2002} Lopez-Corredoira, M., Cabrera-Lavers, A., Garzon, F., \& Hammersley, P.L.\ 2002, \aap, 394, 883
\bibitem[Majewski et al.(2003)]{Majewski2003} Majewski, S.~R., Skrutskie, M.~F., Weinberg, M.~D., \& Ostheimer, J.~C.\ 2003, \apj, 599, 1082
\bibitem[Majewski et al.(2004)]{Majewski2004} Majewski, S.~R., Ostheimer, J.~C., Rocha-Pinto, H.~J., Patterson, R.~J., Guhathakurta, P., \& Reitzel, D.\ 2004, \apj, 615, 738
\bibitem[Martin et al.(2004)]{Martin2004} Martin, N.~F., Ibata, R.~A., Bellazzini, M., Irwin, M.~J., Lewis, G.~F., \& Dehnen, W.\ 2004, \mnras, 348, 12
\bibitem[Momany et al.(2006)]{Momany2006} Momany, Y., Zaggia, S., Gilmore, G., Piotto, G., Carraro, G., Bedin, L.R., \& de Angeli, F.\ 2006, \aap, 451, 515
\bibitem[Newberg et al.(2002)]{Newberg2002} Newberg, H.~J., et al.\ 2002, \apj, 569, 245
\bibitem[Nishiyama et al.(2005)]{Nishiyama2005} Nishiyama, S., et al.\ 2005, \apjl, 621, L105
\bibitem[Ojha et al.(1996)]{Ojha1996} Ojha, D.~K., Bienayme, O., Robin, A.~C., Creze, M., \& Mohan, V.\ 1996, \aap, 311, 456
\bibitem[Ojha et al.(1999)]{Ojha1999} Ojha, D.~K., Bienaym{\'e}, O., Mohan, V., \& Robin, A.~C.\ 1999, \aap, 351, 945
\bibitem[Ojha(2001)]{Ojha2001} Ojha, D.K.\ 2001, \mnras, 322, 426
\bibitem[Phleps et al.(2000)]{Phleps2000} Phleps, S., Meisenheimer, K., Fuchs, B., \& Wolf, C.\ 2000, \aap, 356, 108
\bibitem[Phleps et al.(2005)]{Phleps2005} Phleps, S., Drepper, S., Meisenheimer, K., \& Fuchs, B.\ 2005, \aap, 443, 929
\bibitem[Perryman et al.(2001)]{Perryman2001} Perryman, M.~A.~C., et al.\ 2001, \aap, 369, 339
\bibitem[Reid(1993)]{Reid1993} Reid, M.~J.\ 1993, \araa, 31, 345
\bibitem[Reid \& Majewski(1993)]{ReidMajewski1993} Reid, N., \& Majewski, S.~R.\ 1993, \apj, 409, 635
\bibitem[Reyl{\'e} et al.(2009)]{Reyle2009} Reyl{\'e}, C., Marshall, D.J., Robin, A.C., \& Schultheis, M.\ 2009, \aap, 495, 819
\bibitem[Robin \& Creze(1986)]{Robin1986} Robin, A., \& Creze, M.\ 1986, \aap, 157, 71
\bibitem[Robin et al.(1996)]{Robin1996} Robin, A.~C., Haywood, M., Creze, M., Ojha, D.~K., \& Bienayme, O.\ 1996, \aap, 305, 125
\bibitem[Robin et al.(2000)]{Robin2000} Robin, A.~C., Reyl{\'e}, C., \& Cr{\'e}z{\'e}, M.\ 2000, \aap, 359, 103
\bibitem[Robin et al.(2003)]{Robin2003} Robin, A.~C., Reyl{\'e}, C., Derri{\`e}re, S., \& Picaud, S.\ 2003, \aap, 409, 523
\bibitem[Rocha-Pinto et al.(2003)]{Rocha2003} Rocha-Pinto, H.~J., Majewski, S.~R., Skrutskie, M.~F., \& Crane, J.~D.\ 2003, \apjl, 594, L115
\bibitem[Rocha-Pinto et al.(2004)]{Rocha2004} Rocha-Pinto, H.~J., Majewski, S.~R., Skrutskie, M.~F., Crane, J.~D., \& Patterson, R.~J.\ 2004, \apj, 615, 732
\bibitem[Schlegel et al.(1998)]{Schlegel1998} Schlegel, D.J., Finkbeiner, D.P., \& Davis, M.\ 1998, \apj, 500, 525
\bibitem[Siegel et al.(2002)]{Siegel2002} Siegel, M.~H., Majewski, S.~R., Reid, I.~N., \& Thompson, I.~B.\ 2002, \apj, 578, 151
\bibitem[Skrutskie et al.(2006)]{Skrutskie2006} Skrutskie M.F., et al.\ 2006, \aj, 131, 1163
\bibitem[Tritton \& Morton(1984)]{Tritton1984} Tritton, K.~P., \& Morton, D.~C.\ 1984, \mnras, 209, 429
\bibitem[van Loon et al.(2003)]{vanLoon2003} van Loon, J.~T., et al.\ 2003, \mnras, 338, 857
\bibitem[von Hippel \& Bothun(1993)]{vonHippel1993} von Hippel, T., \& Bothun, G.~D.\ 1993, \apj, 407, 115
\bibitem[Yamagata \& Yoshii(1992)]{Yamagata1992} Yamagata, T., \& Yoshii, Y.\ 1992, \aj, 103, 117
\bibitem[Yaz \& Karaali(2010)]{Yaz2010} Yaz, E., \& Karaali, S.\ 2010, \na, 15, 234
\bibitem[York et al.(2000)]{York2000} York, D.~G., et al.\ 2000, \aj, 120, 1579
\bibitem[Yoshii(1982)]{Yoshii1982} Yoshii, Y.\ 1982, \pasj, 34, 365
\bibitem[Yoshii et al.(1987)]{Yoshii1987} Yoshii, Y., Ishida, K., \& Stobie, R.~S.\ 1987, \aj, 93, 323
\bibitem[Wainscoat et al.(1992)]{Wainscoat1992} Wainscoat, R.~J., Cohen, M., Volk, K., Walker, H.~J., \& Schwartz, D.~E.\ 1992, \apjs, 83, 111
\end{thebibliography}
\end{document}